\documentclass[a4paper,aps,preprint,superscriptaddress,prb]{revtex4}
\pdfoutput=1
\usepackage{revsymb}
\usepackage{amsmath,amssymb}
\usepackage{bm}
\usepackage{fancyhdr}
\usepackage{amsfonts}
\usepackage{amsthm}
\usepackage{booktabs} 
\usepackage{ifthen}
\usepackage[english]{babel}
\usepackage{setspace} 

\usepackage{ifpdf}

\ifpdf
\usepackage[pdftex, a4paper=false, b5paper=true,
       pdftitle={Angular redistribution of near-infrared spontaneous emission from quantum dots in 3D photonic crystals, B.H. Husken},
       pdfauthor={B.H. Husken},
       pdfsubject={Angular redistribution of near-infrared spontaneous emission from quantum dots in 3D photonic crystals},
       bookmarks,
       breaklinks=true,
       colorlinks=true,
       linkcolor=black, 
       anchorcolor=black, 
       citecolor=black,
       bookmarksnumbered=true,
       pagebackref=false,
       pdfstartview=FitH,
       bookmarksopen=false]{hyperref}   

\usepackage[pdftex]{graphicx,color} 
\fi

\usepackage{epsfig,graphicx,color}

\fussy

\begin{document}
\selectlanguage{english}

\title{Angular redistribution of near-infrared emission from quantum dots in 3D photonic crystals}

\author{B.\:H.~Husken} \email{husken@amolf.nl}
\homepage{www.photonicbandgaps.com} \affiliation{Center for
Nanophotonics, FOM Institute for Atomic and Molecular Physics
(AMOLF), P.O. Box 41883, 1009~DB~Amsterdam, The Netherlands.}
\affiliation{Complex Photonic Systems (COPS), MESA$^+$ Institute for
Nanotechnology and Department of Science and Technology, University
of Twente, P.O. Box 217, 7500~AE~Enschede, The Netherlands.}

\author{A.\:F.~Koenderink}
\affiliation{Center for Nanophotonics, FOM Institute for Atomic and Molecular Physics (AMOLF), P.O. Box 41883, 1009~DB~Amsterdam, The Netherlands.}

\author{W.\:L.~Vos}
\affiliation{Center for Nanophotonics, FOM Institute for Atomic and
Molecular Physics (AMOLF), P.O. Box 41883, 1009~DB~Amsterdam, The
Netherlands.} \affiliation{Complex Photonic Systems (COPS), MESA$^+$
Institute for Nanotechnology and Department of Science and
Technology, University of Twente, P.O. Box 217, 7500~AE~Enschede,
The Netherlands.}

\begin{abstract}
We study the angle-resolved spontaneous emission of near-infrared light sources in 3D photonic crystals over a wavelength range from 1200 to 1550~nm. To this end PbSe quantum dots are used as light sources inside titania inverse opal photonic crystals. Strong deviations from the Lambertian emission profile are observed. An attenuation of 60~\% is observed in the angle dependent radiant flux emitted from the samples due to photonic stop bands. At angles that correspond to the edges of the stop band the emitted flux is increased by up to 34~\%. This increase is explained by the redistribution of Bragg-diffracted light over the available escape angles. The results are quantitatively explained by an expanded escape-function model. This model is based on diffusion theory and adapted to photonic crystals using band structure calculations. Our results are the first angular redistributions and escape functions measured at near-infrared, including telecom, wavelengths. In addition, this is the first time for this model to be applied to describe emission from samples that are optically thick for the excitation light and relatively thin for the photoluminesence light.
\end{abstract}

\maketitle

\section{Introduction} \label{sec.intro}
There is much interest in nanophotonic control over the spontaneous emission of light~\cite{Yablonovitch1987aa,Vahala2003aa,Noda2007aa}. Therefore, emission properties of light sources like atoms, molecules, and quantum dots embedded in myriad nanostructures are studied intensively. The spontaneous emission rate of these light sources is described by Fermi's golden rule. The golden rule shows that the spontaneous emission rate is proportional to a local property called the Local Density of Optical States (LDOS)~\cite{Sprik1996aa}. Indeed, recent experiments have demonstrated that it is possible to control the LDOS, and thereby the spontaneous emission rate, using photonic crystals~\cite{Yablonovitch1987aa,Sprik1996aa,Lodahl2004aa,Fujita2005aa,Nikolaev2007aa,Vallee2007aa,Nikolaev2008aa,Julsgaard2008aa}.

A second important aspect of photonic crystals is their ability to strongly influence the directional emission spectrum from embedded light sources. These spectral changes are caused by optical Bragg diffraction~\cite{Yamasaki1998aa,Petrov1998aa,Megens1999aa,Megens1999ab,Schriemer2001aa,Lin2002aa,Koenderink2002aa,de_Dood2003ab,Bechger2005aa,Nikolaev2005aa,Barth2005aa,Li2007ae,Noh2008aa,Brzezinski2008aa}. Although there are many qualitative studies on light transport in photonic crystals, there exist only a few quantitative papers. One of the central concerns regarding light emission from photonic crystals is the angular distribution, or \emph{escape function} of the emitted light over exit angles. Fundamental insight in this escape probability distribution is needed to explain the photonic crystal properties~\cite{Koenderink2005ab,Kaas2008aa}, and the crystals' influence on the radiative emission of the light source. Surprisingly, the escape function is often disregarded, which may easily lead to misinterpretations of experimental results. For example, changes in the emission spectrum, measured from low-efficiency light sources inside photonic crystals may mistakenly be attributed to changes in the LDOS, which cause an enhancement or inhibition of the sources' spontaneous emission power\footnote{For efficient light sources the spontaneous emission rate should be derived from time resolved experiments. For low-efficiency light sources changes in the spontaneous emission rate can be inferred from intensity spectra, see Reference~\cite{Koenderink2002aa}.}. Hence, detailed knowledge about the escape function and a comparison to the emission spectra from non-photonic crystals are needed to find the cause of changes in the measured emission spectrum~\cite{Koenderink2002aa,Koenderink2003ab}.

Most experimental work on light emission in photonic crystals and the corresponding escape function is limited to the visible range. For experiments on very strongly interacting photonic crystals made of semiconductors such as silicon~\cite{Woldering2008aa,Woldering2008ab}, near-infrared emitting light sources, in particular versatile quantum dots such as PbSe and PbS ($\lambda>1100$~nm) are important to avoid absorption$^,$\footnote{Fabrication of strong photonic crystals requires a large refractive index contrast. As silicon has a high refractive index the material is very useful for photonic crystal fabrication.}. The ability to control spontaneous emission and light propagation at these frequencies is of keen interest for applications, as spontaneous emission is a limiting factor in numerous devices~\cite{Yablonovitch1987aa}. Emission control therefore implies a substantial increase in efficiency in applications such as light beaming for the telecom industry, reflector design, and solar cell light control~\cite{Gratzel2001aa}. Furthermore, guiding of light can be used to enhance the collection efficiency in single photon on demand applications. In this paper, emitted light escaping from 3D photonic crystals is studied at near-infrared wavelengths in the range of 1200 to 1550 nm. To this end colloidal PbSe quantum dots are used as light sources inside strongly-photonic titania inverse-opal photonic crystals with a range of lattice parameters.
\section{Experimental details}
\label{sec.experimental}
The samples consist of commercially available PbSe quantum dots (Evident Technologies) embedded inside titania inverse opal photonic crystals. See References~\cite{Wijnhoven1998aa,Wijnhoven2001aa} for a detailed description of the fabrication and characterization of these crystals. X-ray diffraction and scanning electron microscope studies show that the direction perpendicular to the samples' surface corresponds to the (111) direction of the crystal~\cite{Vos1997aa,Wijnhoven2001aa}.

The quantum dots emit at near infrared wavelengths $\lambda\approx1300$~nm, and we estimate the ratio between the lights' transport mean free path $l$ and a typical sample thickness ${\rm L}$ to be $l/{\rm L}\approx 3$ at these wavelengths~\cite{Koenderink2005ab}. Braggs' law was used to derive the required radii $r$ of the polystyrene spheres that yields the crystals' lowest order stop band (which corresponds to the \emph{L-gap} in band-structure calculations) at that emission frequency, taking into account a typical 30~\% shrinking of the crystal during fabrication~\cite{Vos1996aa,Born1997aa,Wijnhoven2001aa}. As a result, polystyrene sphere radii in the range of $r=340~\text{to}~620$~nm are needed. Additionally, $r<200$~nm spheres are used for reference measurements. For these radii, the lattice parameter is too small to cause Bragg diffraction at the quantum dot emission wavelength, and therefore these reference crystals are referred to as \emph{non-photonic}.

Not all fabricated structures are suited for emission experiments. Parts of the crystals may be covered with bulk titania, or the air-sphere ordering may be poor. Visually, samples that show clear Bragg diffraction are selected for further study. Subsequently, bright-field microscope images are made to chart each sample. Finally, reflectivity measurements are used to reveal the areas with clear stop bands that are suitable for subsequent emission experiments.

The useful samples are infiltrated with colloidal PbSe quantum dots; similar to References~\cite{Schriemer2001aa,Koenderink2002aa,Lodahl2004aa,Nikolaev2005aa}. Each sample is placed in a 5~ml glass vial. Next, 50~$\mu$l of 0.92~$\mu$M PbSe quantum dots in hexane suspension is added. After one day the suspension is removed and the samples are rinsed three times with hexane, each time for 20 s. Subsequently, the sample is left to dry for one day. The infiltration leads to a distribution of quantum dots on the titania surfaces inside the inverse opals. From the concentration of dots in the suspension and the volume of the air spheres with typical radii of 385~nm, the distance between adjacent quantum dots is estimated to be 120~nm. This is sufficiently large to avoid reabsorption and energy-transfer processes between the dots~\cite{Lakowicz1999aa}. In the results section below it is deduced that the quantum dots are really inside the photonic crystal.

\begin{figure}[!tbp]
\begin{center}
\includegraphics[width=1.0\columnwidth]{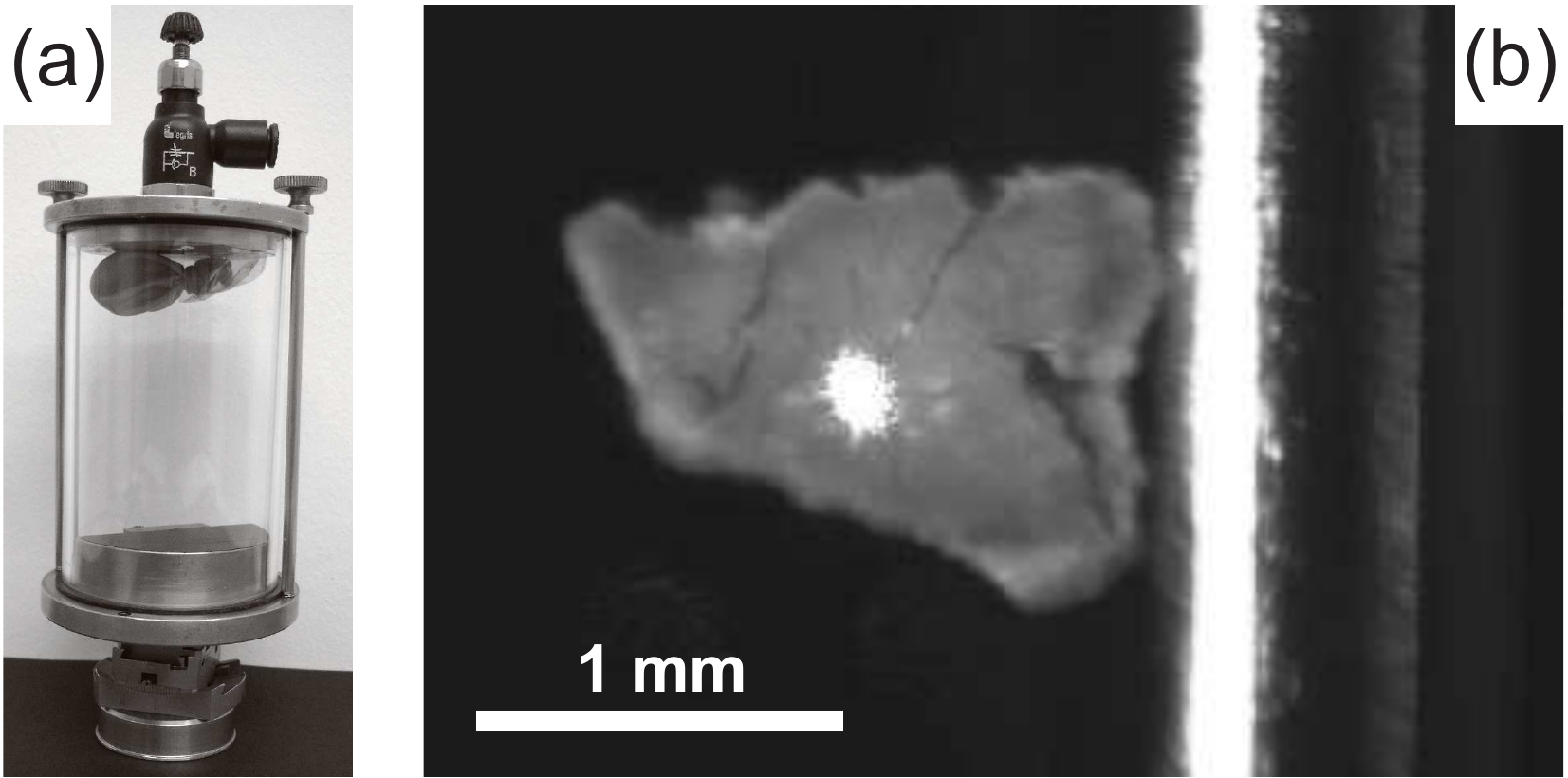}
\caption{(a) Photograph of the sealed chamber used in the optical experiments. A transparent cylinder is sealed off by two plates. The bottom plate is mounted on four stages which are fixed to a rotation stage. The top plate contains a valve used to purge the chamber with nitrogen. (b) Front-view, gray-scale image of a titania inverse opal photonic crystal glued to a needle-pin (on the right). The white spot on the sample is a $\lambda=532$~nm laser spot that is used to locally excite PbSe quantum dots inside this structure. Due to a combination of diffusion, scattering and camera exposure time, the size of the laser spot appears much larger than the size of the actual focus ($\text{radius}\approx7\mu$m).}
\label{fig:sample_on_needle}
\end{center}
\end{figure}

To prevent photo-oxidation of the quantum dots, the sample preparation and handling was carried out in a high quality nitrogen-purged glove box (MBraun, LabStar). For optical measurements the samples were glued to the side of a needle pin, mounted in a sealed chamber, and kept under a 1.6~bar nitrogen atmosphere. Figure~\ref{fig:sample_on_needle}.(a)
shows the home-build chamber that can be mounted on a rotation stage via a threaded ring at the bottom. Between this ring and the chamber, stages are visible that can be used to tilt and translate the chamber. Figure~\ref{fig:sample_on_needle}.(b) contains an image of a sample glued to a needle pin. As only the thin side of the crystal is in contact with the needle, both the frontside and backside of the sample remain accessible for the optical experiments, discussed below.

\begin{figure}[!tb]
\begin{center}
\includegraphics[width=1.0\columnwidth]{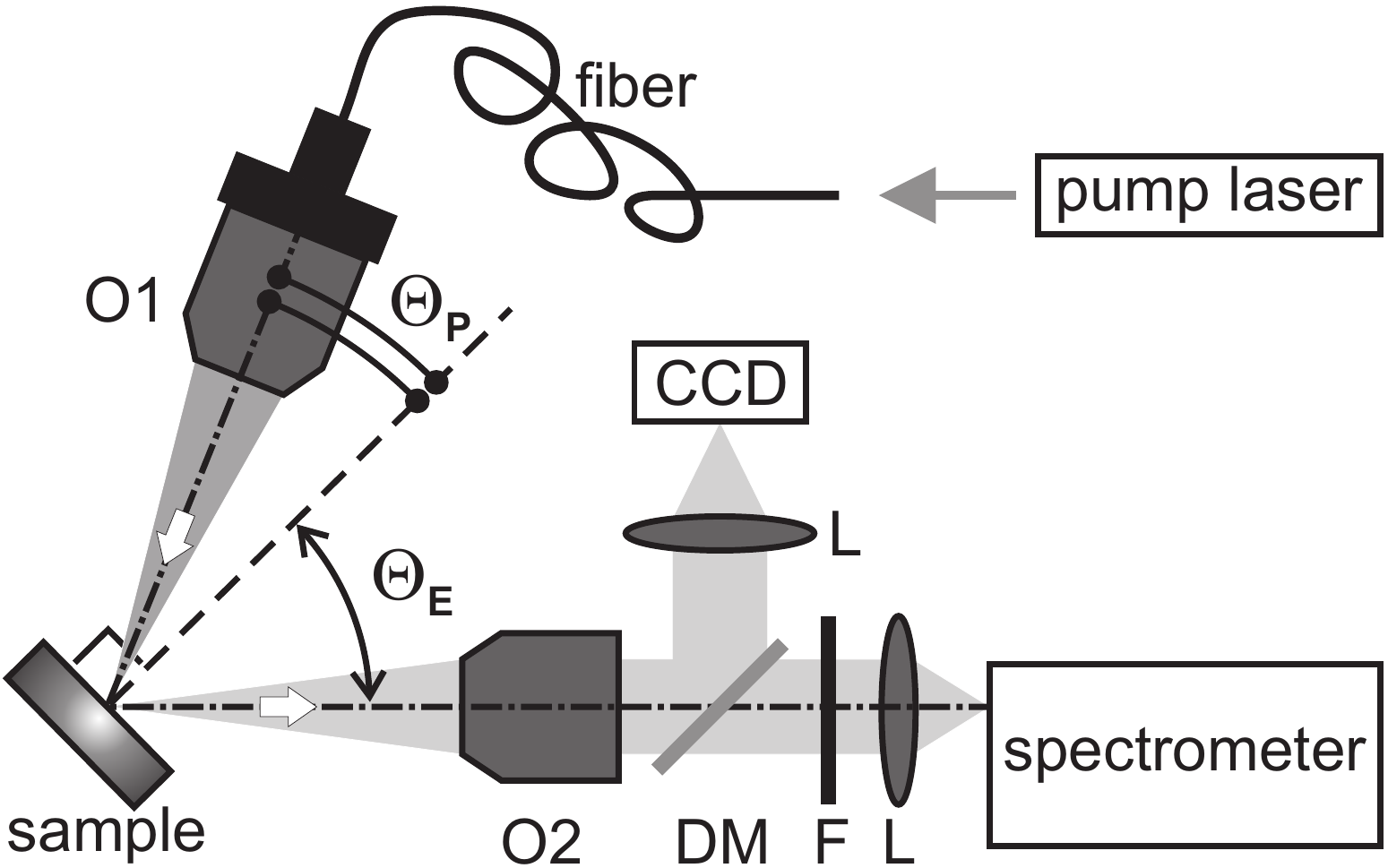}
\caption{Schematic representation of the experiment. Light from a $\lambda=532$~nm pump laser is coupled into a single mode fiber and used to excite quantum dots in a sample, through objective O1. Near-infrared emission from the dots is collected with objective O2 and sent to a spectrometer. A longpass filter (F) blocks the laser light. An achromat lens (L) focuses the collected light onto the entrance slit of the spectrometer. An additional dichroic mirror (DM) is used to guide the visible light to a camera for alignment purposes. The sample is rotated together with the pump beam such that angle $\Theta_{\rm P}$ is fixed and the emission collection angle $\Theta_{\rm E}$ changes. Consequently, the excitation conditions are the same for every angle $\Theta_{\rm P}$. The arrows indicate the propagation direction of the light.}
\label{fig:define_angles}
\end{center}
\end{figure}

Figure~\ref{fig:define_angles} describes the basic concept of the experiment.
The sample is illuminated by a slightly focused laser beam of wavelength $\lambda=532$~nm. The excitation spot radius is estimated to be about 7~$\mu$m. Light emitted from the sample is collected by a 0.05~NA objective and sent to a spectrometer. An 850~nm longpass filter blocks the laser light in the detection path. The sample and excitation objective are rotated together, such that only the detection angle $\Theta_{\rm E}$ changes, as in Reference~\cite{Nikolaev2005aa}. After changing this detection angle a CCD camera is used to roughly reposition the sample such that the illuminated spot is in focus with the spectrometer. Subsequent XYZ fine-tuning of the sample alignment was done by optimization to the emission signal measured with the spectrometer. The sample surface projected onto the entrance slit of the spectrometer changes with $\Theta_{\rm E}$. The measured intensities are corrected for this\footnote{In this paper we adopt the commonly used term \emph{intensity} for what is actually the radiant flux.}, as discussed in Reference~\cite{Nikolaev2005aa}. This correction also takes into account a phenomenological description of the excitation power distribution around the focal spot of the excitation beam. The current configuration allows the emission detection over an unmatched range of both positive and negative angles.

Emission from the titania backbone was experimentally determined to be negligible at the emission wavelength of the quantum dots, whereas backbone emission did exert influence on the measured signals in earlier experiments at visible wavelengths~\cite{Lodahl2004aa}. Reasonably, the titania emission is expected from defect luminescence in the middle of the titania band gap, \emph{i.e.}, in the visible, and not around wavelengths $\lambda=1300$~nm.

To reduce systematic errors, the emission detection angle $\Theta_{\rm E}$ is varied in the following way. First, $\Theta_{\rm E}$ is increased from 0$^\circ$ to maximum positive angles in steps of 20$^\circ$. Subsequently, the angle is decreased in steps of 20$^\circ$ to maximum negative angles, which results in a data set with steps $\Delta\Theta_{\rm E}=10^\circ$. For some measurements the procedure is repeated to obtain a data set with steps $\Delta\Theta_{\rm E}=5^\circ$. The experiment ends where it started, at $\Theta_{\rm E}=0^\circ$. At the start and end of the experiment the excitation laser is blocked after which the background signal is determined. As a result, the reproducibility of the measurement can be judged for all frequencies, and in all experiments. For the typical integration times of 30 to 100~s, error bars of 1~\% of the experimental value can be used to estimate the uncertainty in the measurement. However, this uncertainty does not account for systematic errors caused by spectral changes that may occur for long excitation times, or alignment issues. When the emission spectrum changes with time, saw-tooth-like shapes appear if the emission intensity is plotted versus the external angle $\Theta_{\rm E}$. Therefore, the additional uncertainty in the results can be evaluated after a measurement.

\section{Results and discussion}
\label{sec:angle_resolved_emission_results and discussion}
\subsection{Angle resolved emission}
\label{sec.angle_resolved_emission}
\begin{figure}[!tbp]
\begin{center}
\includegraphics[width=1.0\columnwidth]{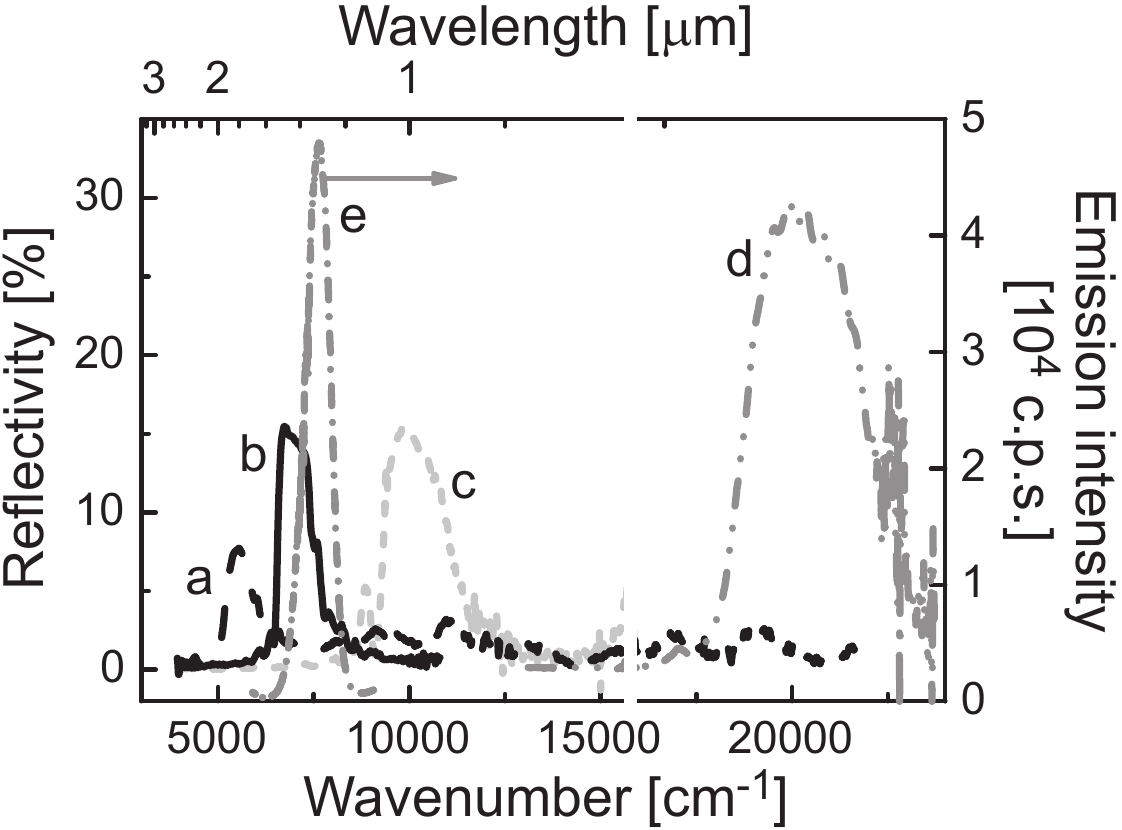}
\caption{Normal-incidence reflectivity spectra measured from four titania inverse opals. These samples were grown from templates with different sphere radii $r_{\rm sphere}$, namely: (a) 655~nm, (b) 515~nm, (c) 403~nm, and (d) 180~nm (left ordinate). These radii correspond with lattice parameters $a$ of 1351~nm (a), 1114~nm (b), 764~nm (c), and 367~nm (d). Stop bands due to the L-gap appear as peaks in the reflectivity spectrum. (e) Shows the emission spectrum of PbSe quantum dots in a hexane suspension (right ordinate). The reflectivity is not defined in a narrow band around 15,800 cm$^{-1}$ because of interference by a reference laser beam inside the spectrometer.}
\label{fig:reflectivity_titania_lubos}
\end{center}
\end{figure}

Reflectivity measurements on four samples with different lattice parameters are shown in Figure~\ref{fig:reflectivity_titania_lubos}. The reflectivity setup was described in References~\cite{Thijssen1999aa,Hartsuiker2008aa}.
Only measurements that show clear Bragg diffraction are shown. The emission spectrum of PbSe quantum dots is included in the same graph. The stop bands are overlapping or to the red of the quantum dot emission spectrum, and far to the blue of the emission spectrum for reference measurements. Therefore, the fabricated samples cover the desired wavelength range. The quality of the selected samples is comparable to samples used in earlier experiments in the visible range~\cite{Koenderink2002aa,Lodahl2004aa,Nikolaev2005aa}. The areas that show the most powerful Bragg diffraction are used in subsequent emission experiments.

\begin{figure}[!tbp]
\begin{center}
\includegraphics[width=1.0\columnwidth,height=0.7\textheight,keepaspectratio=true]{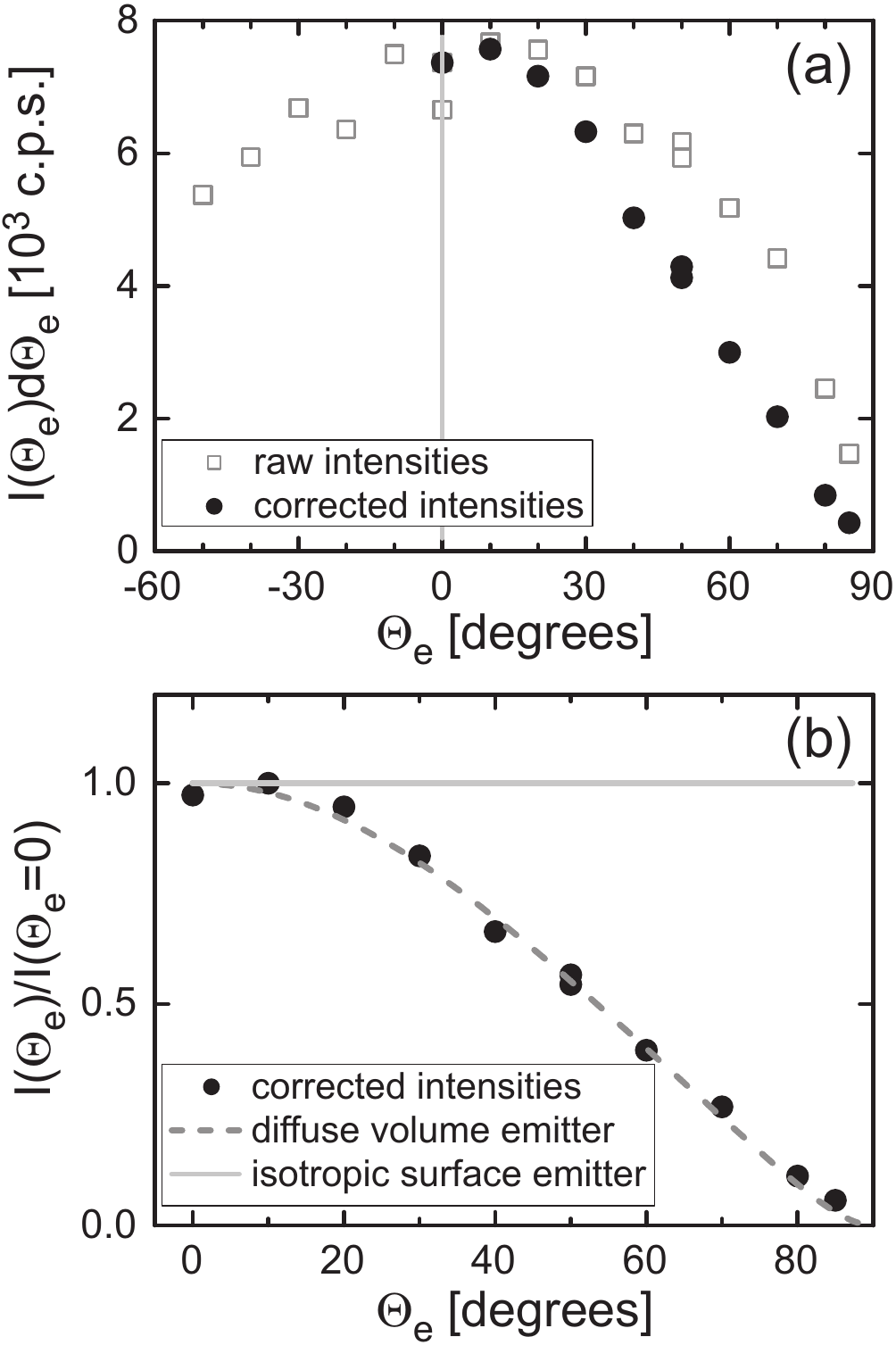}
\caption{Angle-resolved emission intensity of PbSe quantum dots in a non-photonic titania inverse-opal. The lattice parameter $a=637$~nm and the measurement is done at $\lambda=1400$~nm. (a) Open squares are the measured, background subtracted, intensities. Filled dots show the measured data corrected for the detected area. The intensity tends to zero as the angle approaches $\Theta_{\rm e}=90^\circ$. (b) Normalized emission intensity versus detection angle. Filled dots are the same as in (a). The dashed line is the expectation for a diffuse volume emitter. Solid line shows the expected behavior for an isotropic surface emitter. Clearly, our measurements are in correspondence with the diffusion model.}
\label{fig:reference_angle_resolved_emission}
\end{center}
\end{figure}

To verify that the quantum dots are really inside the photonic crystal, a non-photonic sample was tested. Figure~\ref{fig:reference_angle_resolved_emission}.(a)
shows the result of the angle resolved emission experiment. The background subtracted data are shown as open squares. Subsequently each data point is corrected for the angle dependent collection efficiency (black dots). The signal is normalized to its maximum value. The measured intensity profile is the same for positive and negative angles, which is an indication of the diffusive nature of the emitted light. In Figure~\ref{fig:reference_angle_resolved_emission}.(b) the data are compared to models for diffuse volume emitters and isotropic surface emitters. The former model is discussed in Section~\ref{sec.escape_function} below, whereas the latter model yields an angle-independent, constant radiant. The data clearly agree with the volume emitter model. This means that the measured quantum dots are really inside the photonic crystal. The contribution from emitters at the samples' surface can be neglected, as the intensity clearly tends to zero at $\Theta_{\rm E}=90^\circ$.

\begin{figure*}[!tbp]
\begin{center}
\includegraphics[width=1.0\textwidth]{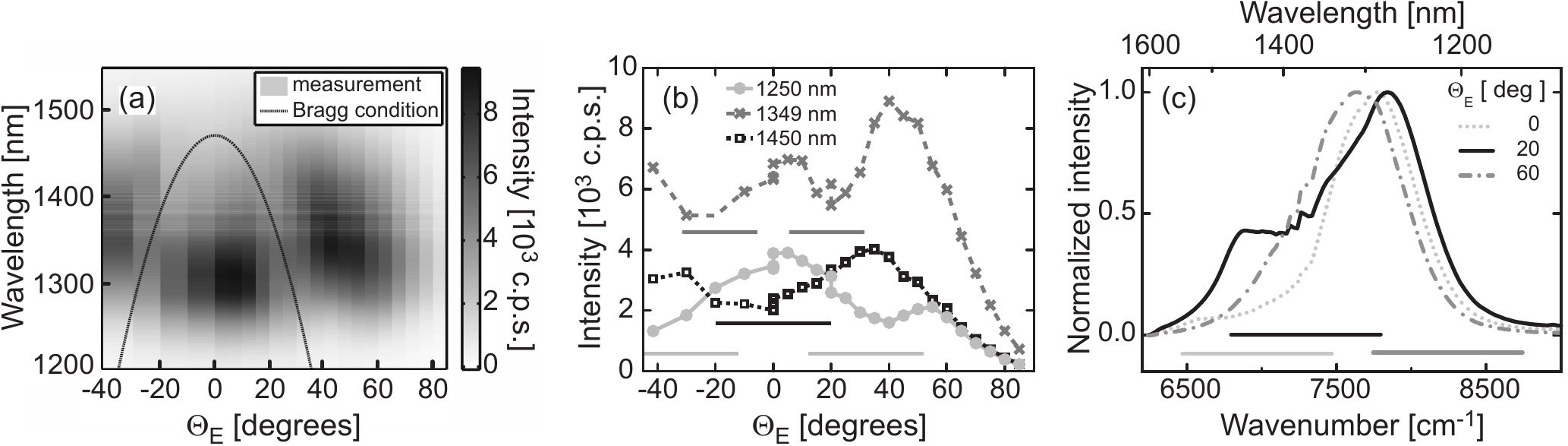}
\caption{Measured, background subtracted and surface corrected PbSe quantum dot emission spectra from a photonic titania inverse opal with lattice parameter $a=1114$~nm. (a) Emission intensity as a function of wavelength and measured angle $\Theta_{\rm E}$ for a photonic sample. A clear suppression of emission intensity is observed. As an estimate for the wavelength of this attenuation the Bragg condition is plotted (black curve). The white areas at wavelengths $\lambda=1200$ and 1550~nm mark the boundaries of the quantum dot emission spectrum. (b) Cross sections of (a) at specific wavelengths shows clear signs of stop bands. (c) Cross sections of (a) at specific angles. The measured emission spectra are angle dependent due to Bragg attenuation of the light in the stop band of the crystal and diffuse scattering. In (b,c) stop band frequencies are indicated by the horizontal lines below the curves. In this experiment spectral changes with time were negligible.}
\label{fig:angle_dep_emission_spectrum_raw}
\end{center}
\end{figure*}

The quantum dot emission is determined to be from the bulk of the crystal. Furthermore, the distribution of the light emitted from a non-photonic crystal shows the expected diffuse emitter profile. These two prerequisites being satisfied, we can start to the use photonic crystals and continue the emission experiments. Figure~\ref{fig:angle_dep_emission_spectrum_raw}
shows extensive data sets measured on a photonic sample. Figure~\ref{fig:angle_dep_emission_spectrum_raw}.(a) Shows the angle and wavelength dependent emission power. A stop band is observed where the emission power is reduced. Increase of detection angle $\Theta_{\rm E}$ results in a shift of the stop band towards shorter wavelengths. To a first approximation this behavior can be explained by Braggs' law (black line)~\cite{Vos1996aa}. In Figures~\ref{fig:angle_dep_emission_spectrum_raw}.(b,c) cross sections of (a) measured at constant wavelengths or angle show the angular and wavelength dependence of the emitted power. The photonic sample behaves very distinctly from a Lambertian emitter. The stop band causes a decrease of the measured emission intensity over an angular range of approximately 40$^\circ$. In (c) the frequency of the emission intensity peak changes with angle. The peak is shifted to the blue if the stop band overlaps the red side of the spectrum, \emph{i.e.}, at small angles. At larger angles the stop band has moved to the blue part of the spectrum and the emission peak is shifted somewhat to the red (not shown). For the largest angle the peak shifts back to the frequency of the dots' actual emission maximum. This experiment demonstrates the influence of the photonic crystals stop bands on the emission spectrum of the quantum dots. This stop band attenuation of a part of the emission spectrum resembles experimental findings using dye molecules~\cite{Yamasaki1998aa,Petrov1998aa,Megens1999aa,Megens1999ab,Schriemer2001aa,Lin2002aa,Koenderink2002aa,Bechger2005aa,Nikolaev2005aa,Barth2005aa,Noh2008aa}, but was not observed with quantum dots before~\cite{Nikolaev2005aa}. An expanded escape function model is used to quantitatively interpret the measurement. This is the topic of the next section.

\subsection{Escape function model}
\label{sec.escape_function}
The model used to interpret the measured emission intensities is based on diffuse light transmission through multiple scattering, opaque media~\cite{Ishimaru1978aa,van_Albada1985aa,Sheng1990aa,Lagendijk1989aa,Zhu1991aa,Vera1996aa}. A complete description is given in References~\cite{Koenderink2003aa,Koenderink2005ab,Nikolaev2005aa}. Here, the most important relations are summarized to explain the essence of this model.

Consider light emission from a sample. The emitted intensity $I(\omega;\mu_{\rm e})$ that escapes the sample at external angles between $\Theta_{\rm E}=\cos^{-1}(\mu_{\rm e})$ and $\cos^{-1}(\mu_{\rm e}+{\rm d}\mu_{\rm e})$ with respect to the surface normal of the sample is written as
\begin{equation} \label{eq:intensity_escape_formula}
I(\omega;\mu_{\rm e}){\rm d}\mu_{\rm e}=I_{\rm tot}(\omega)P(\omega;\mu_{\rm e}){\rm d}\mu_{\rm e}.
\end{equation}
Here, $I_{\rm tot}(\omega)$ that exits the sample through the detection face is the spontaneously emitted intensity at frequency $\omega$. $P(\omega;\mu_{\rm e})$ is a normalized probability distribution function that describes the distribution of photons over the available escape angles. We call $P(\omega;\mu_{\rm e})$ the \emph{escape function} and use it to compare the model to experimental data. To determine $P(\omega;\mu_{\rm e})$ experimentally Equation~\ref{eq:intensity_escape_formula} is rewritten such that
\begin{equation} \label{eq:P_experimental}
P(\omega;\mu_{\rm e})=\frac{I(\omega;\mu_{\rm e})}{I_{\rm tot}(\omega)}.
\end{equation}
The numerator is given by the measured intensity and the denominator $I_{\rm tot}(\omega)$ is determined by summing the measured $I(\omega;\mu_{\rm e})$ weighted by $2\pi \sin(\Theta_{\rm E}){\rm d}\Theta_{\rm E}$ to approximate the integration over 2$\pi$ solid angle.

For diffuse wave transmission through random media the escape function is analytically derived, and given by~\cite{Zhu1991aa,Durian1994aa}
\begin{equation}\label{eq:P_model}
P(\omega;\mu_{\rm e})=\frac{3}{2}\mu_{\rm e}[\tau_{\rm e}(\omega)+\mu_{\rm i}]\cdot[1-R_{\rm D}(\omega;\mu_{\rm i})].
\end{equation}
Here, $\cos^{-1}(\mu_{\rm i})$ is the angle inside the sample, which will be related to $\cos^{-1}(\mu_{\rm e})$ using Snell's law~\cite{Born1997aa}. $R_{\rm D}(\omega;\mu_{\rm i})$ is the angle and frequency dependent internal-reflection coefficient. Finally, $\tau_{\rm e}(\omega)=z_{\rm e}(\omega)/l(\omega)$ is the extrapolation length ratio, defined by the extrapolation length $z_{\rm e}(\omega)$ and the transport mean free path $l(\omega)$ of the sample. $z_{\rm e}(\omega)$ is the distance from the samples' surface, at which the diffuse intensity extrapolates to 0. $\tau_{\rm e}(\omega)$ can be expressed as a function of the angle-averaged reflectivity of the sample boundaries $\bar{R}_{\rm D}(\omega)$~\cite{Lagendijk1989aa,Zhu1991aa,Vera1996aa}, \emph{i.e.},
\begin{equation}\label{eq:extrapolation_length_ratio}
\tau_{\rm e}(\omega)=\frac{2}{3}\Bigg{[}\frac{1+\bar{R}_{\rm D}(\omega)}{1-\bar{R}_{\rm D}(\omega)}\Bigg{]}.
\end{equation}
Furthermore, it was shown how to evaluate this average reflectivity $\bar{R}_{\rm D}(\omega)$ from the angle and frequency dependent internal reflection coefficient $R_{\rm D}(\omega;\mu_{\rm i})$~\cite{Zhu1991aa,Durian1994aa,Vera1996aa}. This results in
\begin{eqnarray}
\bar{R}_{\rm D}(\omega)&=&\frac{3C_2(\omega)+2C_1(\omega)}{3C_2(\omega)-2C_1(\omega)+2}\text{,\hspace{10pt} with}\label{eq:averageR_vs_angle_dep_R}\\
C_n(\omega)&=&\int_0^1 R_{\rm D}(\omega;\mu_{\rm i})~\mu_{\rm i}^n~d\mu_{\rm i}.
\end{eqnarray}

Hence, the escape function from Equation~\ref{eq:P_model} can be evaluated if $R_{\rm D}(\omega;\mu_{\rm i})$ is known. So far the derivation of the model was general and derived for random photonic media in which light transport is diffuse due to multiple scattering. This isotropic model for $P(\omega;\mu_{\rm e})$ successfully described experimental findings on random media when $R_{\rm D}(\omega;\mu_{\rm i})$ was modeled using Fresnel's law in combination with an effective refractive index. However, this does not contain the stop bands observed in photonic crystals. Hence, a different approach is needed to find an expression for $R_{\rm D}(\omega;\mu_{\rm i})$ and expand the escape-function model to describe photonic crystals, a procedure pioneered by our group~\cite{Koenderink2003aa}.

The reflectivity from titania inverse opals is mainly determined by simultaneous Bragg diffraction from (111) and (200) planes~\cite{van_Driel2000aa}. The internal reflection coefficient is therefore modeled as the sum of two Gaussian reflection peaks~\cite{Nikolaev2005aa}, \emph{i.e.},
\begin{eqnarray}\label{eq:R_internal}
R_{\rm D}(\omega;\mu_{\rm i})&=&R_1(\mu_{\rm i})\exp\Bigg{\{}-\frac{[\omega-\omega_1(\mu_{\rm i})]^2}{2\Delta\omega_1(\mu_{\rm i})^2}\Bigg{\}}\nonumber\\
&&{}+R_2(\mu_{\rm i})\exp\Bigg{\{}-\frac{[\omega-\omega_2(\mu_{\rm i})]^2}{2\Delta\omega_2(\mu_{\rm i})^2}\Bigg{\}},
\end{eqnarray}
with angle-dependent peak reflectivities $R_{\rm 1,2}(\mu_{\rm i})$, widths $\Delta\omega_{\rm 1,2}(\mu_{\rm i})$, and center frequencies $\omega_{\rm 1,2}(\mu_{\rm i})$\footnote{For these measurements the parameters used at $\mu_{\rm i}=1$ are: $R_{\rm 1}(1)=0.7$, $R_{\rm 2}(1)=0.3$, $\omega_{\rm 1}(1)=0.797$, $\omega_{\rm 2}(1)=1.33$, $\Delta\omega_{\rm 1}(1)=0.064$, $\Delta\omega_{\rm 2}(1)=0.071$.}. These center frequencies cannot be described by simple Bragg diffraction by 111 and 200 planes, since the two diffraction conditions cross. At this crossing a large avoided crossing occurs~\cite{van_Driel2000aa}. Therefore, band structure calculations are used to model $\omega_{\rm 1,2}(\mu_{\rm i})$~\cite{Koenderink2003aa}. For $R_{\rm 1,2}(\mu_{\rm i})$ and $\Delta\omega_{\rm 1,2}(\mu_{\rm i})$ normal incidence reflectivity data are used. Furthe\begin{scriptsize}\begin{footnotesize}\begin{small}\begin{footnotesize}\end{footnotesize}\end{small}\end{footnotesize}\end{scriptsize}rmore, it is assumed that these four parameters are smooth functions of $\mu_{\rm i}$. This expanded escape-function model can now be used to describe the results on photonic crystals.

\subsection{Comparison experiment with escape function model}
\label{sec.compare_experiment_with_escape_function}
\begin{figure}[!tbp]
\begin{center}
\includegraphics[width=1.0\columnwidth]{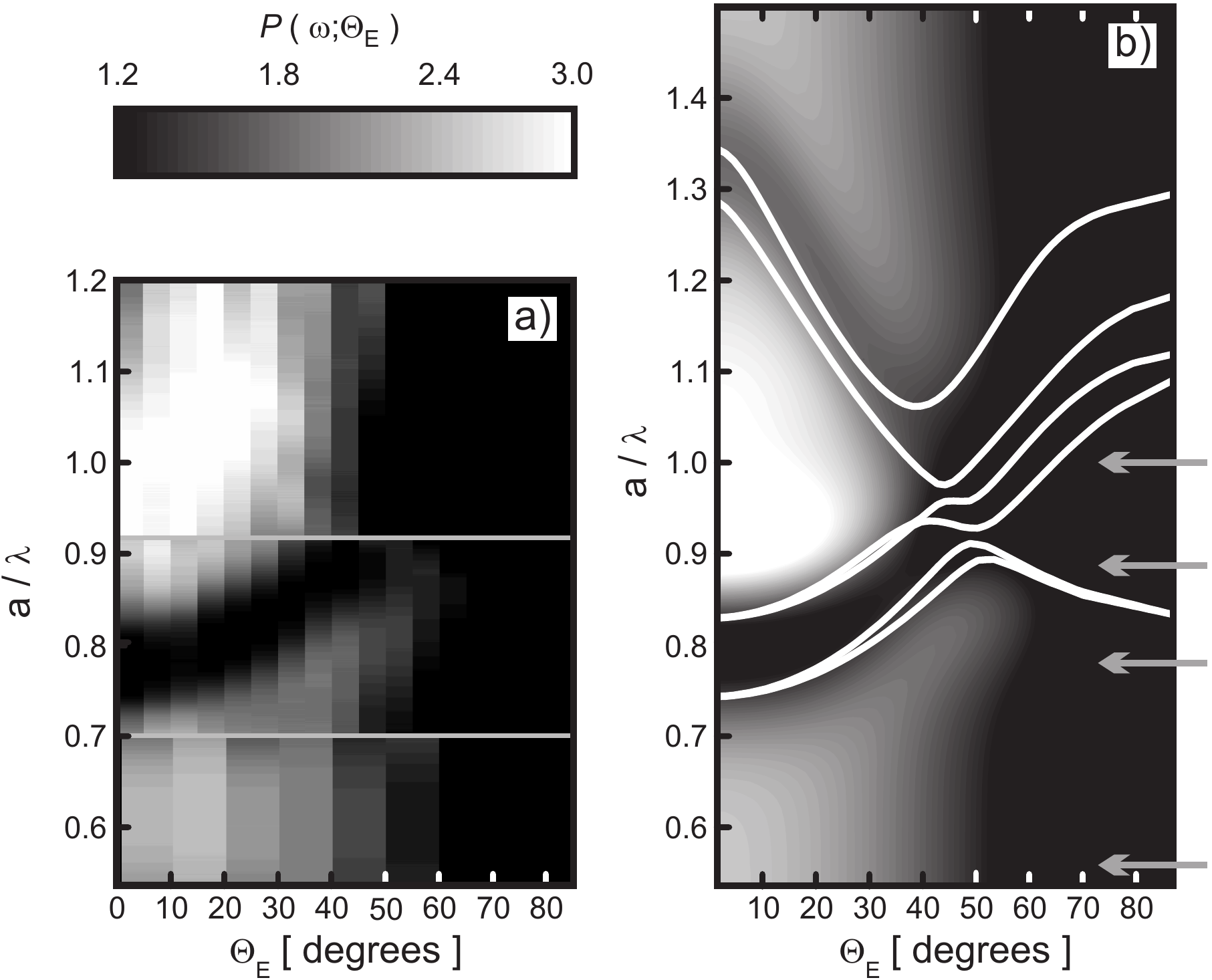}
\caption{Escape probability $P(\omega;\Theta_{\rm E})$. (a) Experimental results from three data sets with different lattice parameters $a$, separated by horizontal lines. Top: $a=1351$~nm, middle: $a=1103$~nm, and bottom: $a=830$~nm were determined from reflectivity measurements and assuming an average refractive index $n_{eff}=1.155$. The low-frequency region shows the expected, Lambertian-like behavior. The central frequency region shows clear inhibition of the escape probability due to Bragg diffraction. Regions of enhanced escape probability are observed around $a/\lambda=0.85 \text{ and } 1.0$ for $\Theta_{\rm E}\approx 50^\circ \text{ and } 0^\circ$ respectively. (b) Result from the expanded escape-function model. For $a/\lambda<1.2$ the model is in agreement with the experimental findings. The stop band caused by Bragg diffraction from (111) lattice planes is observed, together with the regions of enhanced escape-probability. The higher order Bragg wave diffraction is more pronounced in the model. It is recognized by the inhibited escape probability region that starts at $a/\lambda=1.3$ for $\Theta_{\rm E}=0$ and continues to lower frequencies at larger angles. White curves show the lowest 6 bands along the LU direction, using parallel momentum ($k_{\|}$) conservation at the surface to transform internal propagation angles into external propagation angles. The horizontal arrows indicate the frequencies of the cross sections used in Figure~\ref{fig:P_comparison} below. The common gray-scale range is limited for clarity.}
\label{fig:calculated_P}
\end{center}
\end{figure}

Figure~\ref{fig:calculated_P}
shows the escape function $P(\omega;\mu_{\rm e})$ over large frequency ranges, derived from experiments and calculated. Figure~\ref{fig:calculated_P}.(a) is derived from three experiments on samples with different lattice parameters. The results from the different experiments are separated by the horizontal, gray lines. The central part ($0.7<a/\lambda<0.92$) shows a strong stop band in the escape probability due to the L-gap. As expected, the frequency of the stop band increases with external angle. An enhanced escape probability is observed at $a/\lambda\approx 0.84$ for $\Theta_{\rm E}\approx 50^\circ$. The enhancement of the escape probability stems from a redistribution of the light that is Bragg diffracted inside the photonic crystal, as discussed below. At lower frequencies ($a/\lambda<0.7$) there is no sign of a stop band and the emission escape probability decreases continuously with $\Theta_{\rm E}$. At high frequencies ($a/\lambda>0.92$) another region is observed with enhanced escape probability at small $\Theta_{\rm E}$. The escape probability rapidly decreases at larger angles. Dark areas at the largest frequencies point out the presence of inhibited escape probabilities caused by Bragg-wave-coupling from the (200) lattice planes~\cite{Koenderink2003aa}. For applications this map can be used to find the lattice parameter $a$ that yields the desired angular dependent escape probability at the wavelength of choice.

Figure~\ref{fig:calculated_P}.(b) shows the calculation from the expanded escape-function model. The calculated escape probability shows the same characteristics as the experimental results. There is a clear inhibition of escape probability starting at $a/\lambda\approx0.8$, for $\Theta_{\rm E}=0^\circ$. The central frequency of this inhibition shows the same dependence on $\Theta_{\rm E}$ as the experimental results. Another line of inhibited escape probability runs from $a/\lambda=1.33$ at $\Theta_{\rm E}=0^\circ$ towards $a/\lambda=0.95$ at $\Theta_{\rm E}=50^\circ$. Multiple Bragg-wave-coupling causes the escape probability to be inhibited for angles larger than 40$^\circ$ and frequencies $0.88<a/\lambda<1.03$. This result agrees well with our experimental results. The regions with the measured, enhanced escape probability where also predicted by the model. Clearly, the expanded escape-function adequately describes the experimental data measured from these strongly photonic crystals. The white curves show the lowest 6 LU bands from a band structure calculation used to describe experiments in the visible regime~\cite{Koenderink2003aa}. All band energies are multiplied by 1.125 such that the lowest bands overlap with the measured L-gap edges. The multiplication factor does strikingly not vary per sample. The difference between experiments and band structure calculations are subject to further study: more information on the titania distribution in these large lattice parameter crystals is needed to verify the model that describes the unit cell. Furthermore, dispersion causes the refractive index to be different in the near-infrared wavelength regime. Nevertheless, this first comparison already shows a very good overlap between the calculated band structure and the escape function.

\begin{figure}[!tbp]
\begin{center}
\includegraphics[width=1.0\columnwidth,height=0.7\textheight,keepaspectratio=true]{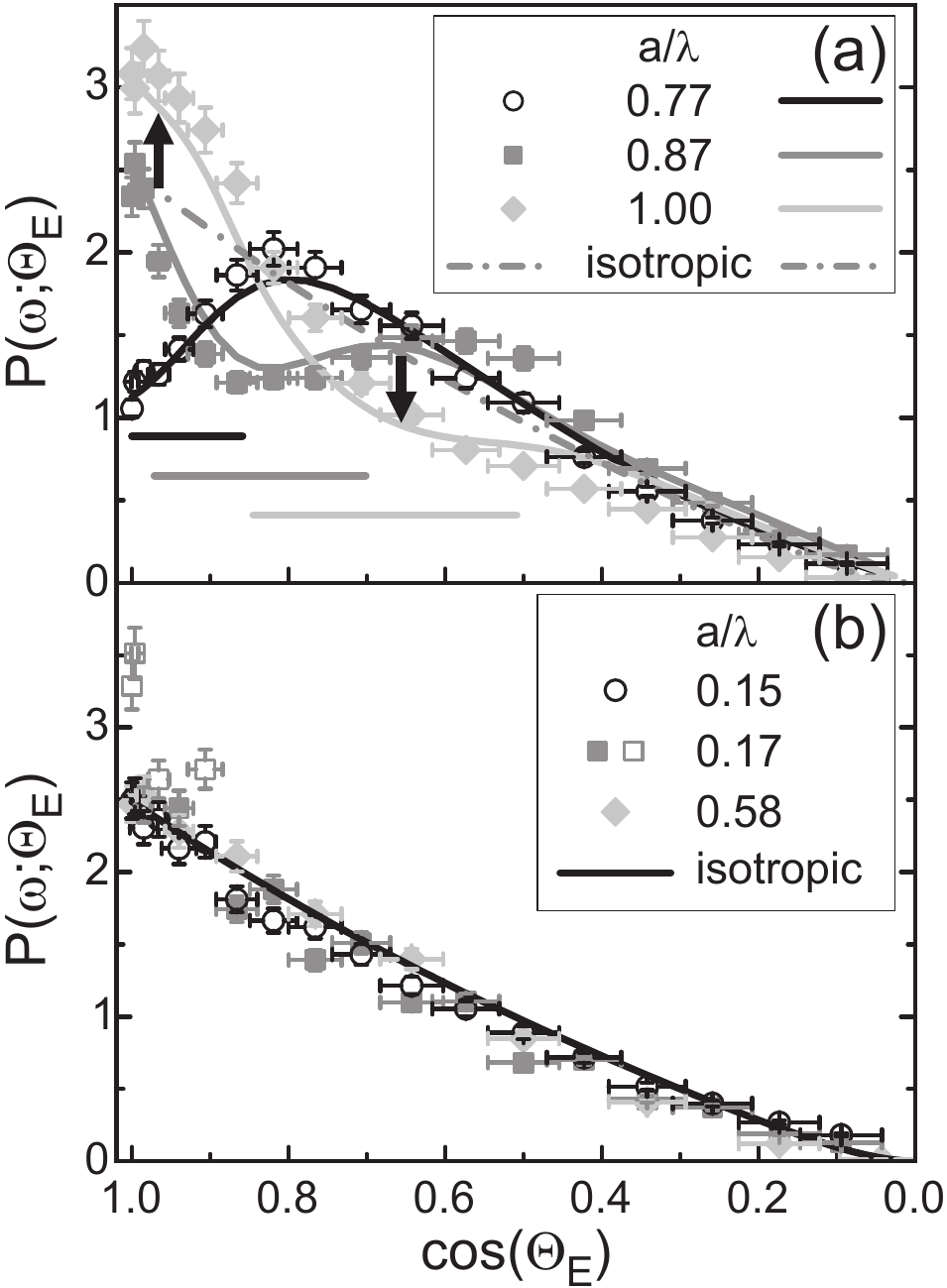}
\caption{Cross sections of escape function at different reduced frequencies $a/\lambda$. Symbols denote the measurements and lines stem from the escape function models. (a) Results in photonic range. The solid lines result from the expanded escape-function model, and the dash-dotted line results from the regular escape-function model. Escape function versus $\cos(\Theta_{\rm E})$ shows inhibition and enhancement of the escape probability with respect to the isotropic distribution (dash-dotted line). Inside (outside) the stop bands the escape probability is inhibited (enhanced) up to 60~\% (34~\%) with respect to the isotropic distribution. The frequency of the stop band is marked by the horizontal lines below the curves. The stop band moves to higher frequencies at larger angles. For the  $a/\lambda=1.00$ case, the upward and downward pointing arrows respectively indicate the enhanced and inhibited escape probability with respect to the isotropic model. (b) Results in non-photonic regime. Circles and squares are measured on the same sample with lattice parameter $a=220$~nm. The diamonds are data collected from a sample with $a=830$~nm. Only for $a/\lambda=0.17$ the experimental data deviate from the model (open squares), as a result of an emission intensity increase with time.}
\label{fig:P_comparison}
\end{center}
\end{figure}

To compare the results from the model and the experiments in more detail, cross sections were made at the frequencies indicated by the arrows in Figure~\ref{fig:calculated_P}.(b). In contrast to Figure~\ref{fig:calculated_P}, the escape function is now plotted versus $\mu_{\rm e}=\cos(\Theta_{\rm E})$. In this representation the escape function for isotropic emitters tends towards a line, simplifying comparison with experimental results. Figure~\ref{fig:P_comparison}.(a)%
shows the escape function versus external angle at three different reduced frequencies in the photonic regime ($a/\lambda\in\{0.77,0.87,1.0\}$). The stop band is clearly visible as the region where the escape probability is reduced with respect to the isotropic probability distribution. Outside the stop band the escape probability is higher than the isotropic model such that the escape probability is conserved. This probability enhancement is caused by a redistribution of the light inside the photonic crystal: for a certain angular range the light Bragg diffracts back into the sample, where it becomes diffuse and escapes the sample at angles away from the Bragg condition. The enhancement is most profound close to the stop band. Further away from the stop band the data converge to the isotropic case. The expanded escape-function model agrees very well with the experimental data. The dash-dotted line shows the regular escape-function model for isotropic emitters. The experimental data appear even more explicit than the model predicts. Compared to isotropic emission, inhibition of the emitted power as high as 60~\% is observed in the photonic crystal. Furthermore, the photonic crystal gave rise to emission power enhancements as high as 34~\%. For $a/\lambda=0.77$ and~0.87 the inhibition and enhancement is observed over a cone with a top angle of $2\cdot\Theta_{\rm E}=60^{\circ}$, which corresponds to a solid angle of $2\pi(1-\cos(\Theta_{\rm E}))=0.84$~steradian. The pronounced inhibition at $a/\lambda=0.87$, between $\cos(\Theta_{\rm E})=0.97~\text{to}~0.71$, corresponds to a solid angle of 1.6 steradian.

Figure~\ref{fig:P_comparison}.(b) shows the escape function $P(\omega;\mu_{\rm e})$ cross sections measured from two non-photonic samples. Symbols denote measured data and the line is calculated from the regular, isotropic model. There is no stop band. The data at $a/\lambda=0.15\text{ and }0.17$ are measured from the same sample. The data at $a/\lambda=0.15\text{ and }0.58$ match the regular escape-function model. The $a/\lambda=0.17$ data deviate from this model, \emph{i.e.}, especially at small angles ($\cos(\Theta_{\rm E})\approx 1$, open squares) the measured data are above the isotropic model. This deviation is explained by an observed emission intensity increase with time, at the blue side of the emission spectrum ($a/\lambda=0.17$). This intensity increase results in a saw-tooth shaped measured angle-resolved emission spectrum, because of the applied order in the measured angles. At $\Theta_{\rm E}=0^\circ$ the result of the very first and last measurement was averaged. Due to this intensity increase the normalization used to calculate ${I_{\rm tot}(\omega)}$ in Equation~\ref{eq:P_experimental} becomes less accurate. This causes the measurement to deviate from the model. The modulation of the measured escape function with external angle is much smaller at the red side of the spectrum ($a/\lambda=0.15$). Although the blue side of the emission spectrum may have changed with time the rest of the spectrum has remained practically constant. This clearly illustrates the effect of inhomogeneous broadening in the quantum dot ensemble, \emph{i.e.}, the measurements at various wavelengths probe different subsets of quantum dots. Clearly, the regular escape-function model fits the escape-function data from non-photonic titania inverse opals very well over a very large range of reduced frequencies. Hence, we are able to model the escape function both from photonic and non-photonic crystals, using our new expression for the internal-reflection coefficient $R_D(\omega;\mu_{\rm i})$ (Equation~\ref{eq:R_internal}).

\subsection{Discussion}\label{sec.discussion}
Photonic crystals can be divided into (1) optically thick samples where the thickness ${\rm L}$ is much larger than the transport mean free path $l$ ($l/{\rm L}\ll 1$), and (2) optically thin samples where the thickness is about the mean free path of light, or less ($l/{\rm L}\gtrapprox 1$). Inside optically thick samples light that is Bragg diffracted back into the sample, scatters off defects and becomes diffuse. Subsequently, this light can exit the structure at different angles. This results in the redistribution of the emission intensity over escape angles. In emission measurements this redistribution causes an emission intensity increase at the edges of the stop band that is clearly not related to effects in the local density of optical states; see also~\cite{Koenderink2003aa,Blum2008aa}. In optically thin samples most emitted light escapes the sample ballistically. There is only little scattering in these structures. Bragg diffraction, however, may still decrease the emission intensity considerably in specific directions~\cite{Megens1999aa}.

The samples used in this study have $l/{\rm L}\approx 3$, which means that for the quantum dot emission the sample is in the intermediate regime between being optically thick ($l\ll{\rm L}$) and beeing optically thin ($l\gtrapprox{\rm L}$). As a consequence a significant part of the light emitted by the dots need not become diffuse before escaping the photonic crystal. Still our model which is based on diffusion theory describes the experiments very well. Although the diffuse nature of the emitted light may be questioned, it is believed that the models' requirement for a distributed light source is fulfilled as (1) the excitation of the quantum dots is diffuse because the sample is optically thick for the 532~nm excitation light, (2) the absorption length of the light is about two orders larger than the sample thickness, due to the low concentration of quantum dots, and (3) the emission of the quantum dots randomizes the wavevector $\vec{\bf{k}}$.

\begin{figure}[!tbp]
\begin{center}
\includegraphics[width=0.8\columnwidth,height=0.7\textheight,keepaspectratio=true]{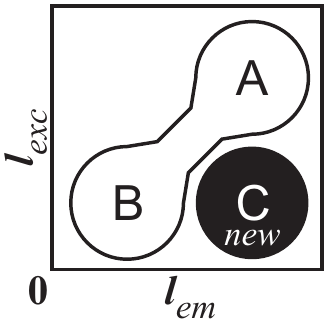}
\caption{Schematic representation of different regimes in photoluminesence experiments. The regimes are defined by the transport mean fee path of the emitted light $l_{em}$ and the excitation light $l_{exc}$, with respect to the sample thickness ${\rm L}$. The regimes marked with A and B have been studied before. A new regime C is defined where the excitation light is diffuse and the emitted light is on the edge of being diffuse or ballistic. The escape function model, based on diffusion theory, was succesfully applied in this new regime.}
\label{fig:l_regimes}
\end{center}
\end{figure}

The good correspondence between the model and the experimental findings demonstrate the applicability of the escape function model to a new regime in photoluminesence experiments. This regime is illustrated in Figure~\ref{fig:l_regimes} 
and defined by the transport mean free paths of the emitted light ($l_{em}$) and the excitation light  ($l_{exc}$), both with respect to the sample thickness ${\rm L}$. In Figure~\ref{fig:l_regimes} the regime marked by "A", where $l_{em},l_{exc}\gg{\rm L}$, corresponds to optically thin samples. Regime "B", with  $l_{em},l_{exc}\ll{\rm L}$, corresponds to optically thick samples. In these regimes the escape function has been used succesfully. In this paper a new regime "C" is added with $l_{exc}\ll{\rm L}$ and $l_{em}\approx{\rm L}$. As discussed above, the escape funtion model, based on diffusion theory and expanded to photonic crystals, was succesfully applied in this new regime.

Strong photoluminescence modifications may easily lead to misinterpretation of experimental findings. There exist various examples where an attenuation in a small frequency range of the photoluminescence emission spectrum is explained by a change in the spontaneous emission properties of the emitter~\cite{Blanco1998aa,Yamasaki1998aa,Yoshino1998aa,Romanov2000aa,Lin2002aa}. In contrast, the attenuation is caused by Bragg diffraction that causes a non-isotropic escape probability. This effect can be understood from Equation~\ref{eq:intensity_escape_formula}: the emitters properties and the local density of states are contained in $I_{\rm tot}(\omega)$, whereas the experimentalist measures $I(\omega;\mu_{\rm e})$. In photonic crystals these spectra cannot be linked without considering the escape function $P(\omega;\mu_{\rm e})$. Without the escape function it cannot be recognized if light is redistributed over directions that do not show Bragg diffraction. Naturally, lifetime measurements could solve the matter, but were not addressed in the articles mentioned above.

The escape function from titania inverse opals was also studied using externally injected light~\cite{Koenderink2003aa}, and, in the visible, using embedded CdSe quantum dot emitters~\cite{Nikolaev2005aa}. A strong redistribution of the light over exit angles was observed, similar to our results. An interesting difference is that in the CdSe quantum dot emission spectra, no stop bands or spectral changes were observed, even though the relative linewidth $\Delta\lambda/\lambda=0.05$ was comparable to the relative stopband width $\Psi\approx0.1$. In contrast, we observe spectral stop bands, similar to earlier work on dyes~\cite{Yamasaki1998aa,Petrov1998aa,Megens1999aa,Megens1999ab,Schriemer2001aa,Lin2002aa,Koenderink2002aa,Bechger2005aa,Nikolaev2005aa,Barth2005aa,Noh2008aa}, in PbSe emission spectra with a relative linewidth $\Delta\lambda/\lambda=0.147$.
\section{Conclusions}
\label{sec.conclusions}
We have studied the angular distribution of near-infrared spontaneous emission ($1200<\lambda<1550$) from PbSe quantum dots inside 3D titania inverse opals. We have observed angular redistribution of the light over the exit angle, caused by a combination of light diffusion and Bragg diffraction in the photonic crystals. The escape function was extracted from the experimental data and explained with a diffusion model, expanded to photonic crystals. We found a very good agreement between the measurement and the model. This is the first time for the model to be applied to the regime of diffuse excitation in combination with optically rather thin samples for the emitted light. Furthermore, an interpretation of emission enhancement at the sides of the stop bands is given in terms of angular redistribution of emission intensities. In the photonic crystal an escape inhibition of 60~\% is observed as well as an enhancement of 34~\%, both with respect to the escape distribution from non-photonic samples. Typically the emission is enhanced or inhibited over solid angles larger than 0.84~steradian. The data presented form the first experimental evidence of angular redistribution of quantum dot emission from photonic crystals at near-infrared wavelengths, including telecom frequencies.

\section*{Acknowledgments} \label{sec:Acknowledgement}
Charles Uju and Steven Kettelarij are acknowledged for the fabrication of the titania inverse opals. This research was supported by NanoNed, a nanotechnology programme of the Dutch Ministry of Economic Affairs, and by a VICI fellowship from the "Nederlandse Organisatie voor Wetenschappelijk Onderzoek" (NWO) to WLV. This work is also part of the research programme of the "Stichting voor Fundamenteel Onderzoek der Materie" (FOM), which is financially supported by the NWO.

\bibliographystyle{phreport}
\bibliography{bibliography_and_articles/20090302_093851_litdbase_final}

\end{document}